\documentclass[sigconf]{acmart}
\usepackage[ruled,vlined]{algorithm2e}
\usepackage{graphicx} % add this in your preamble
\usepackage{booktabs}
\AtBeginDocument{%
  \providecommand\BibTeX{{%
    \normalfont B\kern-0.5em{\scshape i\kern-0.25em b}\kern-0.8em\TeX}}}

\setcopyright{acmlicensed}
\copyrightyear{2025}
\acmYear{2025}
\acmDOI{XXXXXXX.XXXXXXX}

\acmConference[HT '25]{In Proceedings of the 36th ACM
Conference on Hypertext and Social Media}{September 15--19,
  2025}{Chicago, IL}
\acmISBN{978-1-4503-XXXX-X/18/06}

\graphicspath{{./images/}}

\begin{document}

%%\title{eScooter-BERT: A Transformer Model for e-Scooter Maneuver Classification and Contextual Embedding}
\title{HyperSumm-RL: A Dialogue Summarization Framework for Modeling Leadership Perception in Social Robots}

\author{Subasish Das}
\affiliation{%
  \institution{Texas State University}
  \city{San Marcos}
  \country{USA}}
\email{subasish@txstate.edu}


\begin{abstract}

%% older
%%  E-scooter maneuver classification is a critical task in micromobility safety analysis, particularly for understanding crash dynamics and informing safety interventions. This study proposes \textit{eScooter-BERT}, a transformer-based model fine-tuned on crash narratives to classify e-scooter maneuvers based on the Pedestrian and Bicycle Crash Analysis Tool (PBCAT) framework. Additionally, \textit{eScooter2Vec} is introduced as a contextual embedding framework that generates dense vector representations of crash narratives to facilitate downstream tasks such as similarity search and clustering. Experimental evaluation of eScooter-BERT, RoBERTa, and DistilBERT demonstrates that eScooter-BERT achieves superior classification performance with a macro-averaged F1-score of 0.65 and accuracy of 64\%, effectively distinguishing between four maneuver types: Crossing Left (CL), Crossing Right (CR), Parallel Same (PS), and Parallel Opposite (PO). While RoBERTa and DistilBERT provide computational efficiency, they exhibit higher misclassifications in CR and PS categories, indicating limitations in capturing detailed maneuver contexts. eScooter2Vec effectively captures semantic nuances, enhancing contextual representation and supporting unsupervised analysis in crash data analytics. The developed framework underscores the efficacy of transformer-based models in maneuver classification and contextual embedding generation, offering a comprehensive toolset for advancing micromobility safety research.

This paper introduces HyperSumm-RL, a hypertext-aware summarization and interaction analysis framework designed to investigate human perceptions of social robot leadership through long-form dialogue. The system utilizes a structured Natural Language Processing (NLP) workflow that combines transformer-based long dialogue summarization, leadership style modeling, and user response analysis, enabling scalable evaluation of social robots in complex human-robot interaction (HRI) settings. Unlike prior work that focuses on static or task-oriented HRI, HyperSumm-RL captures and hypertextually organizes dynamic conversational exchanges into navigable, semantically rich representations which allows researchers to trace interaction threads, identify influence cues, and analyze leadership framing over time. The contributions of this study are threefold: (1) we present a novel infrastructure for summarizing and linking long, multi-turn dialogues using leadership-style taxonomies; (2) we propose an interactive hypertext model that supports relational navigation across conversational themes, participant responses, and robot behavior modes; and (3) we demonstrate the utility of this system in interpreting participant trust, engagement, and expectation shifts during social robot leadership scenarios. The findings reveal how hypertextual workflows can augment HRI research by enabling transparent, interpretable, and semantically grounded analysis of emergent social dynamics.

\end{abstract}

%%
%% The code below is generated by the tool at: http://dl.acm.org/ccs.cfm



%% The following are not a requirement, delete if not using
%\received{9 May 2025}  %% inital submission date
%\received[revised]{July 14 2025} %% interim new draft
%\received[accepted]{July 14 2025}  %% publication version

\maketitle

\section{Introduction}

In various aspects of daily life, including personal interactions, workplace communication, and online forums, dialogue plays a crucial role. This has garnered significant attention from both the academic and industrial communities \citep{Zhang2020}. The rise of speech recognition and remote work has led to widespread recording of long conversations (e.g., interviews, meetings, and conferences) creating a wealth of information that's difficult to digest quickly. To address this, NLP tasks like dialogue summarization, segmentation, and question answering have been developed to help users extract key insights efficiently. However, discussions on leadership, opinions on leaders, and potential biases are complex and often involve extended dialogues that pose challenges for current NLP systems \cite{Zhong2022}. While recent language models have advanced NLP tasks, their general-purpose training limits their performance in domain-specific, extended dialogues.

Research into social robots and intelligent technologies has gained much traction in recent years. By using other studies findings and addressing gaps in their research we can continue to develop this emerging field. The selection criteria for sources reviewed in this study is that the study must concern social robots intelligence, benefits and applications, and the acceptability. These studies all contribute to incorporation of social robots into society, the role that they play, and public opinion of these new technologies. 

Several studies have explored how social robots exhibit contextual and emotional intelligence by analyzing human responses. Ahn et al. \cite{Ahn2019} emphasized integrating socially intelligent service robots into daily life, focusing on their form, function, behavior, and user expectations. However, deploying such systems requires effective training strategies. Addressing this, Chi et al. \cite{Chi2022} found that humans naturally use mixed teaching methods—such as instruction and evaluation—when training robots, suggesting social learning is more nuanced than previously assumed. To develop emotionally intelligent robots, Doewes \cite{Doewes2024} introduced the SEAI system, combining reactive and symbolic reasoning to simulate human-like conversations and emotional judgments. Evaluating such complexity is challenging, prompting \cite{Cho2023} to propose the Social Robot Intelligence Quotient (SRIQ), which quantifies robot capabilities across five areas: human recognition, user modeling, body movement, self-expression, and communication—each tailored to the robot's morphology. The emotional impact of robot interaction is significant. Spaccatini et al. \cite{Spaccatini2023} found that anthropomorphic design influences empathy, with agency and experience attribution affecting how users relate to others post-interaction. Similarly, Xu et al. \cite{Xu2023} showed that users model behavior based on robots’ social presence and perceived expertise, even without direct outcomes.

In this study, we applied DialogLM, which is a neural encoder-decoder model specifically designed and trained for understanding and summarizing extended dialogues. DialogLM is built upon the sequence-to-sequence model architecture \cite{Zhong2022} and can be used for a large variety of NLP tasks. Built on a sequence-to-sequence architecture, DialogLM is pre-trained using five strategies that disrupt content and speaker order, forcing the model to learn dialogue structure and flow. Its dual attention mechanism—sparse for local details and global for context—allows efficient processing of dialogues over 8,000 words.

\subsection{Motivation of Current Study}

This study explores a key gap in human-robot interaction by evaluating social robots as interview leaders, focusing on how leadership styles affect candidate perceptions. While hiring platforms streamline connections, interviews remain complex and inconsistent. The research introduces a hypertext-based dialogue summarization workflow using large language models to extract key insights from long, multilingual, multi-turn conversations. Framing dialogues as hypertext, the study offers a novel approach to building scalable, interpretable, and socially aware interview systems.

\section{Methodology} \label{sec:method}
\subsection{Datasets}
\textbf{Intrerview Data}: We draw on open-source data from Cichor et al. study \cite{Cichor2023} and focuses on understanding perceptions of social robots as leaders through qualitative analysis. Participants interacted with either a transformational or transactional robot leader, followed by semi-structured interviews lasting about 45 minutes. These interviews captured individual reactions, opinions, and experiences with the robot’s leadership style. The analysis centers on eight in-depth interviews to explore how users perceive social robots in positions of authority.

\textbf{Training Data}: We utilized three widely recognized benchmarks for the tasks mentioned above: AMI \citep{Carletta2005}, ICSI  \citep{Janin2003}, and QMSum \citep{Zhong2022}. AMI and ICSI comprise transcripts of meetings, with AMI focusing on meetings for products and their designs in corporate settings and ICSI on academic group meetings in educational institutions. Each of these datasets includes meeting summaries and meticulously annotated topic boundaries. On the other hand, QMSum serves as a baseline for query-based meeting summarization tasks the span across multiple domains. This dataset incorporates two query types: general queries, which are suitable for summarization tasks, and specific queries, which can be employed for abstractive question-answering tasks. Additionally, QMSum provides human-annotated topic boundaries.

\subsection{Framework}

We propose a structured framework for summarizing interview dialogues, encompassing data preparation, target generation, robustness enhancement, and model evaluation. The process begins with a curated dataset of interview questions and responses, followed by thorough preprocessing and formatting into JSON sequence format compatible with model input requirements. Target summaries are then generated using a pre-trained transformer model (e.g., T5), serving as ground truth for evaluation. To simulate real-world variability, controlled noise is injected into the training data, enhancing model robustness. The DialogLM model is subsequently fine-tuned on this augmented dataset and evaluated on a held-out test set using ROUGE metrics. This workflow enables the generation of accurate, contextually rich summaries, offering a scalable solution for extracting key insights from long-form interview content. Figure 1 displays the overall framework.

\begin{figure}[h!]
    \centering
    \includegraphics[width=\columnwidth]{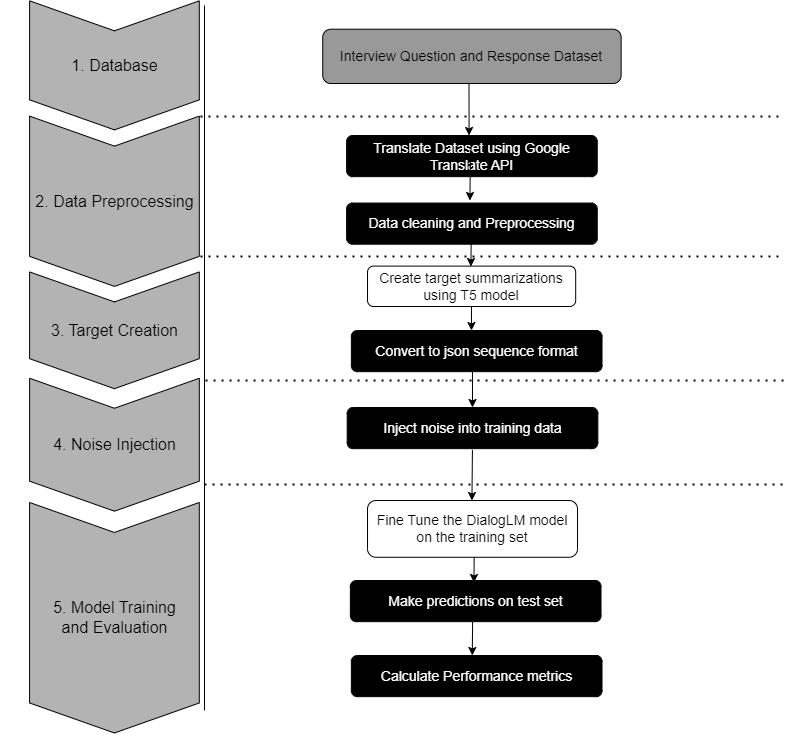}
    \caption{Study framework.}
    \label{fig_dimensions}
\end{figure}

\subsection{Pre-processing}
The original data is in German, so translation is needed to properly use this data with the DialogLM model. The Google Translate API is used to translate text from one language to another \citep{Han2023}. Each row in the DataFrame is iterated over, and both the Question and Response columns are translated from their original language to the desired target language.  After translating the individual question and response, the data is grouped by unique interview number. Consequently, all questions and responses from the same interview are combined into a single group. Within each group, the translated 'question and response' pairs are concatenated into a single text string. The final step of preprocessing involves further cleaning of the text data. This includes the removal of special characters and the handling of words that weren't translated correctly.

%\subsection{Transformer Models}
%\begin{figure}[h]
   % \centering
   % \includegraphics[width=.7\columnwidth]%{DialogLM.drawio.png}
    %\caption{Model architecture for DialogLM.}
   % \label{fig:fig_Arch}
%\end{figure}

\textbf{DialogLM}: Built on the Transformer architecture, DialogLM addresses two limitations of existing models like UNILM: the lack of dialogue-specific pretraining and short input capacity. We introduce a window-based denoising pretraining method that incorporates structured dialogue knowledge. To handle long inputs, we integrate a hybrid attention mechanism—combining sparse Sinkhorn attention with full self-attention in the 4th, 8th, and 12th encoder layers—enabling efficient processing of long dialogues while maintaining strong summarization quality.

\textbf{Text-to-Text Transformer}: T5, a scalable transformer-based model, is employed for summarization. With configurable parameters ranging from 60 million to 11 billion, T5 offers robust performance across various NLP tasks.

\textbf{Target Creation}: Using Hugging Face Transformers, we apply T5 to summarize interview transcripts. Each dialogue turn is prefixed with 'Q:' and 'R:' to maintain structure. Summaries are generated from tokenized inputs, decoded, and stored for downstream use.

\textbf{Noise Insertion}: To improve robustness, we apply five noise strategies during pretraining:

\begin{itemize}
    \item \textbf{Speaker Mask}: 50\% of speaker names are replaced with \texttt{[MASK SPEAKER]}.
    \item \textbf{Turn Splitting}: The longest turn is split into multiple turns; subsequent ones are anonymized.
    \item \textbf{Turn Merging}: Consecutive turns are randomly merged; only the first speaker identity is preserved.
    \item \textbf{Text Infilling}: Random spans are masked using a Poisson distribution.
    \item \textbf{Turn Permutation}: Turns are randomly shuffled to disrupt the dialogue sequence.
\end{itemize}

\subsection{Experimental Settings}
Table \ref{tab:hyperparameters} shows the modeling hyperparameters for all datasets were done on a NVIDIA V100 TENSOR CORE GPU in the Google Colab environment. These settings were chosen to allow for quick testing without consuming all of the available GPU RAM. The limited training steps due to hardware constraints could definitely have an impact on these results, and likely warrant further investigation.

\begin{table}[ht]
\caption{Hyperparameters for Model Training.}
\label{tab:hyperparameters}
\centering
{\fontsize{9}{11}\selectfont
\begin{tabular}{ll}
\toprule
\textbf{Parameter} & \textbf{Value} \\
    \midrule
Label Smoothing Factor & 0.1 \\
Per Device Train Batch Size & 1 \\
Per Device Eval Batch Size & 2 \\
Gradient Accumulation Steps & 1 \\
Max Source Length & 3000 \\
Max Target Length & 360 \\
Learning Rate & 2e-5 \\
Warmup Steps & 50 \\
Max Steps & 10 \\
Save Steps & 5 \\
Eval Steps & 5 \\
\bottomrule
\end{tabular}
}
\end{table}

For tracking the performance of the DialogLM models two main performance metrics are used Rouge-N and Rouge-L.

\textbf{ROUGE N}: ROUGE-N is a metric concerned with the recall of n-grams. Specifically, it assesses the similarity between a candidate summary and a set of reference summaries. The ROUGE-N score is derived through the following equation:

\begin{equation}
ROUGE-N = \frac{{\sum_{gram \in S} \min(Count_{gram}, Count_{gram}^{match})}}{{\sum_{gram \in S} Count_{gram}}}
\end{equation}

The 'n' here represents the length of n-grams, known as $gram_n$.
$Count_{gram}^{match}$ refers to the highest count of n-grams that occur in both the reference summaries and the candidate summary. An n-gram is a contiguous sequence of n items (or words) from a given sample of text or speech. ROUGE-N is recall-oriented, as its denominator reflects the total n-grams in the reference summaries. In contrast, BLEU focuses on precision by measuring the proportion of overlapping n-grams between the candidate and reference translations \citep{Papineni2002}. 

\textbf{ROUGE-L (Longest Common Sequence)}: Consider a sequence $Z$ represented as $[z_1, z_2, \ldots, z_n]$, which is deemed a subsequence of another sequence $X$ denoted as $[x_1, x_2, \ldots, x_m]$ if there exists a strictly increasing sequence $[i_1, i_2, \ldots, i_k]$ of indices from $X$ such that for all $j = 1, 2, \ldots, k$, it holds that $x_{i_j} = z_j$ \citep{Cormen2009}. Now, turning to two sequences, $X$ and $Y$, we define the Longest Common Subsequence (LCS) as the common subsequence of $X$ and $Y$ that possesses the maximum length. In various contexts, LCS has proven invaluable. For instance, it has been employed in identifying potential cognate candidates during the construction of N-best translation lexicons from parallel texts. \cite{Melamed1995} introduced the concept of the LCS Ratio (LCSR), a metric that gauges the degree of cognateness between two words. 

\section{Results}
Table \ref{tab:results} and Table \ref{tab:test-results}  present the evaluation results of a language model trained on the QSUM dataset and tested on the Interviews dataset, along with the AMI and ICSI datasets for comparison. It is important to note that the QSUM dataset was used for training due to its suitability for model development, while predictions were made on the Interviews dataset for which training data was insufficient, containing only eight samples. Table \ref{tab:results} illustrates the performance metrics for the QSUM, AMI, and ICSI datasets. The R-1, R-2, and R-L scores represent the quality of generated text summaries, with higher scores indicating better performance. The Loss metric reflects the model's training loss, where a lower value is generally preferred.

\begin{table}[h]
\centering
\begin{minipage}{0.45\textwidth}
  \centering
  \caption{Evaluation Results}
  \label{tab:results}
  \begin{tabular}{lcccc}
    \toprule
    Dataset & R-1 & R-2 & R-L & Loss \\
    \midrule
    QSUM & 19.22 & 2.64 & 10.66 & 5.78 \\
    AMI & 29.16 & 4.69 & 13.96 & 5.35 \\
    ICSI & 27.77 & 4.10 & 12.04 & 5.91 \\
    \bottomrule
  \end{tabular}
\end{minipage}

\vspace{6pt} 
\begin{minipage}{0.45\textwidth}
  \centering
  \caption{Test Prediction Results}
  \label{tab:test-results}
  \begin{tabular}{lcccc}
    \toprule
    Dataset & R-1 & R-2 & R-L & Loss \\
    \midrule
    Interviews & 15.06 & 7.86 & 26.74 & 4.29 \\
    AMI & 29.76 & 4.08 & 13.55 & 5.29 \\
    ICSI & 27.02 & 2.89 & 12.15 & 5.91 \\
    \bottomrule
  \end{tabular}
\end{minipage}
\end{table}

Table \ref{tab:test-results}, presents the model evaluation in the interviews, AMI, and ICSI datasets, assessing generalization to unseen data. While the Interviews dataset shows slightly lower ROUGE-1 scores, indicating modest unigram prediction performance, it achieves higher ROUGE-2 and ROUGE-L scores, reflecting improved bigram accuracy and coherence. Additionally, lower training loss suggests effective learning of interview-specific patterns. These results highlight the model’s ability to transfer knowledge from the QSUM dataset—chosen for its question-response dialogue structure similar to interviews—demonstrating the value of transfer learning for limited data scenarios. The variation in ROUGE scores across datasets underscores the importance of domain-aligned training data in NLP. Summarization also enhances interview analysis by condensing key content. Below is an example summary from an interview, where (Q) is the interviewer and (R) is the volunteer reflecting on their interaction with the social robot:

\begin{quote}
\leftskip=0.001em % Adjust this value as needed (default is ~2em)

\textbf{Q}: To what extent do you think he can be a role model for his employees?

\textbf{R}: At least to me, it didn’t seem like he had specifically observed me or adjusted his facial expressions and gestures. It felt more like a pre-programmed sequence that was always the same. The appearance matched that impression. The gestures were relatively good but still a bit choppy—not a fluent language. I also had some difficulty with clarity. No, I wouldn’t say he responded directly. Still, he came across as friendly. It was relatively well done. He conveyed things with passion and gave the impression that something was expected from you. As mentioned, his enthusiasm in communication was quite good. The gestures looked relatively human-like.

\textbf{R}: And the voice needs to be worked on a bit more to make it more human. Gestures and yes he doesn’t have facial expressions. But that might also be a step to incorporate.

\end{quote}

\subsection{Content Analysis}
The content analysis revealed distinct patterns in how participants perceived robot leaders, shaped by their underlying leadership style. Rather than emerging in isolated or sequential statements, these perceptions appeared as interconnected reflections, often revisited and reframed throughout the interviews. This observation supports our hypertextual framework, which conceptualizes dialogue not as a linear stream but as a web of semantically linked nodes—emotional cues, behavioral judgments, and expectations that surface repeatedly across different turns.

\textbf{Reaction to robot behavior:}
Participants consistently viewed the transformational robot leader as more motivating and goal-oriented, often describing it as a role model. Its expressive and passionate communication elicited enthusiasm and deeper engagement. However, such impressions were not confined to a single portion of the dialogue—they resurfaced intermittently, layered within discussions about trust, fairness, and leadership qualities. These cross-references underscore the non-linear nature of human reasoning, captured within our hypertext model as links between emotionally and cognitively resonant fragments.

\textbf{Emotional response:} Participants expressed mixed emotional reactions that shifted throughout the conversation. While some appreciated the robot’s emotional neutrality for its perceived efficiency, others voiced discomfort with its inability to express empathy—especially in relation to the transactional robot. These concerns about emotional disconnection did not appear in a single thematic block; rather, they were embedded in discussions about leadership style, communication effectiveness, and personal comfort. The hypertextual framing helps illuminate these recurring and interdependent concerns, showing how emotional reactions to robot behavior are built cumulatively across different dialogue layers.

\textbf{Leadership applicability}: Participants’ views on robot leadership were shaped by a range of multimodal and linguistic factors. While the transformational robot’s friendliness and passion were noted, its credibility was undermined by unnatural tone, scripted gestures, and imprecise language. These issues—though mentioned at various points in the interview—collectively reflect a broader skepticism about the robot’s communicative competence. By interpreting such dispersed insights through a hypertext lens, we identify how impressions about leadership credibility are not formed in isolation but constructed through interlinked reflections on behavior, emotional tone, and human-robot rapport. As communication is central to effective leadership \citep{Gemeda2020}, the absence of contextual flexibility and natural rhythm in the robot’s delivery significantly limited its influence. Participants in team-based scenarios raised further concerns about the robot’s motivational capacity, citing a lack of personal connection. 

\section{CONCLUSIONS}
This study explored how perceptions of robot leaders—transformational and transactional—manifest in interview dialogue and how these conversations can be meaningfully summarized using advanced language models. We implemented DialogLM, enhanced through a window-based denoising method, to manage the complexity of long, speaker-rich interactions.

A central contribution of our work is the conceptualization of interview dialogue as hypertext. Rather than treating dialogue as a linear sequence of utterances, we view it as a network of interconnected semantic units—reflecting shifts in emotion, leadership evaluation, and topic framing. This hypertextual perspective informs both model design and interpretation. DialogLM operates within this structure by linking thematically related segments across speaker turns, enabling it to reconstruct not just what was said, but how meaning evolves throughout the conversation. This framework supports a more robust summarization process, capable of capturing layered human-robot interactions. By aligning summarization techniques with the inherent non-linearity of dialogue, we extend the capabilities of LLMs to better reflect the structure and depth of interview discourse—particularly where social and leadership themes intersect.

Although the model demonstrated strong performance, challenges remain in handling subtle emotional cues, context-switching, and domain-specific content. Public datasets for interview-style dialogues are limited, and reliance on generated summaries restricts evaluation precision. Structural differences between interviews and multi-party meeting data may also affect generalizability.

\end{document}